\documentstyle[aps,preprint,epsf]{revtex}

\def\char{characteristic\ }
\def\la{{\lambda}}
\def\ze{{\zeta}}
\def\zei{{\zeta_i}}
\def\zeo{{\zeta_0}}
\def\sgn{{\rm sgn}}
\def\erf{{\rm erf}}

\def\eps{{\epsilon}}
\def\lao{{\la_0}}

\def\rbg{{\bf R}}

\def\be{{\beta             }}
\def\de{{\delta}}
\def\si{{\sigma}}
\def\La{{\Lambda}}
\def\De{{\Delta            }}

\def\beq{\begin{equation} }
\def\eeq{\end{equation}   }
\def\beqa{\begin{eqnarray} }
\def\eeqa{\end{eqnarray}   }
\def\thl{N\rightarrow\infty}
\def\rel{n\rightarrow0}

\begin{document}

\title{A Variational Approach to the Localization Transition of
Heteropolymers at Interfaces}

\author{ A.~Trovato and A.~Maritan }

\address{International School for Advanced Studies (SISSA),\\
and Istituto Nazionale di Fisica della Materia,\\
Via Beirut 2-4, 34014 Trieste, Italy }

\address{The Abdus Salam International Center for Theoretical Physics,\\
Strada Costiera 11, 34100 Trieste (Italy) }

\maketitle


\begin{abstract}

A chain with random hydrophobic-hydrophilic charges is studied in the
presence of an interface separating a polar from a non polar solvent.
Within a Gaussian variational approach in replica space,
a transition is found, from a high temperature
region, where the chain is delocalized, to a low temperature region,
where the chain is localized at the interface. The transition temperature
diverges as the neutrality of the chain is approached. Our results
are in agreement with an Imry-Ma type argument by Garel
{\it et al}\/. \cite{GHLO}. The problem of replica symmetry breaking
is also addressed, and the replica-symmetric solution is found to be
unstable. 

\vspace{10pt}
PACS numbers: 05.40.+j, 36.20.-r, 87.15.-v
\vspace{10pt}
\end{abstract}

Random heteropolymers are often studied in connection with the
behavior of complex macromolecules such as proteins 
\cite{BryWol,GarOrl,ShaGut}. 
From a theoretical point of view, random heteropolymers are easier to
study, since one can use many tools introduced in the framework of
the statistical mechanics of disordered systems. In the case of
proteins, it is believed \cite{Anf,Dill} that the main force driving
the protein towards the folded state in physiological conditions
is the interaction of the amino-acids with the polar solvent.
In the folded state, hydrophobic (non-polar) residues are buried in
the chain interior, whereas hydrophilic (polar or charged) ones are
predominantly confined at the surface.

Interesting questions arises when an interface separates a
polar and a non-polar solvent. The latter may be thought to model
the lipidic environment inside the cell membrane \cite{Cre}.
It is indeed known that membrane proteins have a higher concentration
of hydrophobic amino-acids than proteins in solution \cite{Cre}.
Moreover, random heteropolymers at interfaces 
have also a technological relevance, due to their effectiveness in the
reinforcing of interfaces between two immiscible polymers
\cite{BrDeGr,DOKHJ}.

In this letter we study, both analytically and numerically, a
very simple model of an ideal random hydrophilic-hydrophobic
chain in presence of an interface separating a polar solvent
(e.g. water) and a non-polar one (e.g. oil). The model has been
firstly introduced by Garel {\it et al}\/\cite{GHLO} (recently, more
specific model for membrane proteins has been introduced \cite{BoSe}). 
In their work it is shown, with Imry-Ma type arguments \cite{IMa} and
some analytical observations, that the quenched randomness in the
sequence of hydrophobic charges causes a localization transition at
sufficiently low temperature. The transition temperature becomes
infinite in the case of a neutral chain, i.e. a neutral chain is
always localized. The case of self-avoiding neutral chains at an
asymmetric selective interface has been also studied with scaling 
arguments \cite{SoDa,SoPeBlu2} and Monte-Carlo simulations
\cite{SoPeBlu,PeSoBlu}, yielding a similar scenario.

We have employed a Gaussian variational approach for the calculation
of the replicated partition function, in the same spirit as in references
\cite{ShaGut2,MePa,MKD}. Within this approach it should be possible to cope
with the possible presence of many metastable states separated by high
energy barriers \cite{Eng}, via the usual Parisi ansatz for replica
symmetry breaking (RSB) \cite{MPV}. On the other hand, the choice of a
quadratic trial Hamiltonian does not provide the exponential decay of the
monomer density which has been found both in homopolymer adsorption at
interfaces \cite{DeGen69} and in Monte Carlo simulations of the model
considered here \cite{PeSoBlu}.
Nevertheless, we think the capability to explore correctly
replica space, typical of the Gaussian approach, broadly makes up for
this shortcoming. This is also justified `a posteriori' by our results.

Within a replica-symmetric ansatz, we 
obtain explicitly the temperature dependent localization
length for a neutral chain. Our result correctly reproduces the expected 
scaling behavior in the limit of high temperatures,
whereas in the limit of low temperatures it shows a non trivial
dependence on the statistical segment length of the chain, the
only microscopic length in the problem. For a non neutral chain we obtain
the expected localization transition, which turns out to be continuous.
Within a one step RSB ansatz, we show the replica-symmetric solution to 
be unstable, which is usually interpreted as a signal of a phase space
decomposition in many energy valleys \cite{MPV}.

\vspace{10pt}

The partition function of an ideal chain in $d$ dimensions \cite{CloJan}, 
in the presence of an interface at $x=-x_0$, is:

\beq
Z\left\{\zei\right\} = {\rm Tr}_{\left\{\rbg_i\right\}}
\exp\left[-{d\over 2l^2}\sum_{i=1}^N\left(\rbg_i-\rbg_{i-1}\right)^2+
\be\sum_{i=0}^N\zei\;\sgn\left(x_i+x_0\right)\right]\: ,
\label{1}
\eeq
where $\be=1\;/\;K_BT$ is the inverse temperature and
$\rbg_i=\left(x_i,\rbg_{i\parallel}\right)$ ($0\le i\le N$) are the positions
of the monomers, $\rbg_{i\parallel}$ being the coordinates parallel to
the interface.
The trace operation is simply a multiple integral over all possible
positions of the $N+1$ monomers.
The chain connectivity is implemented as usual with a harmonic
potential, $l$ being the typical length of the chain bond.
The hydrophobic charge $\zei$ is assigned to the $i$-th monomer,
and determines its interaction energy with the solvent. If the
monomer is on the polar solvent side (say $x>-x_0$), it takes the
energy $-\zei$, whereas if it is on the non-polar solvent side
($x<-x_0$), it takes the energy $\zei$. Thus, when $\zei>0$ ($\zei<0$),
the monomer is hydrophilic (hydrophobic) and prefers the right side
(the left side). The hydrophobic
charges $\zei$ are identically distributed independent random
Gaussian variables, with average $\zeo$ and variance $\De$,
$P\left(\zei\right)=\exp\left[-\left(\zei-\zeo\right)^2/\:2\De^2\right]\:
/\sqrt{2\pi\De^2}\:$; $\zeo=0$ corresponds to a neutral chain.
Any self-interaction between monomers, such as self-avoidance, is
neglected in this simple approach. Hence, integration over
$\rbg_{i\parallel}$ can be carried out immediately and the problem
becomes one-dimensional.

We are interested in the quenched free energy per
monomer $f_q = -1/\left(N\be\right)\:\overline{\ln Z\left\{\zei\right\}}$,
where $\overline{\cdot}$ denotes the average over disorder.
We introduce $n$ replicas of the chain for a given disorder realization
$\left\{\zei\right\}$ in order to employ the usual replica trick:
$f_q = \lim_{\rel}\left(\overline{Z^n}-1\right)/n$.
Thus, we deal with an effective one-dimensional homopolymer hamiltonian
\beqa
{\cal H}_n = & - & {\be^2\De^2\over2}\sum_{i=0}^N\left(\sum_{a=1}^n\;
\sgn(x_{i,a}+x_0)\right)^2 - \nonumber \\
 & - & \sum_{a=1}^n\left[\be\zeo\sum_{i=0}^N\;\sgn\left(x_{i,a}+x_0\right)
 - {d\over 2l^2}\sum_{i=1}^N\left(x_{i,a}-x_{i-1,a}\right)^2\right]\: .
\eeqa

The Gaussian variational approach consists in approximating ${\cal H}_n$
with a trial function ${\cal H}_t$ which is quadratic in the monomer
coordinates $x_{i,a}$, for integer $n$.
This provides the bound $\ln\overline{Z_n}\ge
\ln Z_t+\langle{\cal H}_t-{\cal H}_n\rangle_{{\cal H}_t}$, where
$Z_t={\rm Tr}_{\left\{x_{i,a}\right\}}\exp\left[-{\cal H}_t\right]$.
For simplicity we consider the case of a ring
polymer ($x_{0,a}=x_{N,a}$), in such a way to exploit the translation
invariance along the chain, and we introduce the corresponding Fourier
coordinate $k$ (${\tilde x}_a\left(k\right)=\sum_{i=1}^Nx_{i,a}
\exp\left(-\:{\rm i}\:ik\right)$). In the thermodynamic limit ($N\gg1$)
we get:

\beqa
{\cal H}_t & = & {1\over2}\int_{-\pi}^{\pi}{{\rm d}k\over2\pi}
\sum_{a,b}{\tilde x}_a\left(k\right)\La_{ab}\left(k\right)
{\tilde x}_b^{*}\left(k\right)\: ,\nonumber \\
\La_{ab}\left(k\right) & = & \left[{2d\over l^2}\left(1-\cos k\right)
+\mu\right]\de_{ab}-\si_{ab}\left(1-\de_{ab}\right)\: ,
\label{4}
\eeqa
where the variational parameters $\mu$ and $\si_{ab}$ have been introduced,
associated with single replicas and replica pairs, respectively.

Let us consider now macroscopic quantities such as the local density
of $a$-monomers, $\rho_a\left(x\right)=\sum_i\de\left(x_{i,a}-x\right)$,
and the so called overlap of replicas $a$ and $b$,
$q_{ab}\left(x,x'\right)=\sum_i\de\left(x_{i,a}-x\right)
\de\left(x_{i,b}-x'\right)$, with $a<b$.
Not surprisingly, the quadratic ansatz (\ref{4}) yields a Gaussian
structure for the average of both of them:

\beqa
\left<\rho_a\left(x\right)\right>_{{\cal H}_t}\; & = &
\;N\:\exp\left[-{x^2\over2\lao}\right]/\left(2\pi\lao\right)^{1/2}\: ,
\label{5}\\
\left<q_{ab}\left(x,x'\right)\right>_{{\cal H}_t}\; & = & \;N\:
{\exp\left[-{1\over4}{\left(x+x'\right)^2\over\lao+\la_{ab}}
-{1\over4}{\left(x-x'\right)^2\over\lao-\la_{ab}}\right]\over
2\pi\left[\left(\lao+\la_{ab}\right)\left(\lao-\la_{ab}\right)\right]^{1/2}}
\: ,\label{6}
\eeqa
where $\lao=\int_{-\pi}^{\pi}{\rm d}k\left[\Lambda^{-1}
\left(k\right)\right]_{aa}/2\pi$ and $\la_{ab}=\int_{-\pi}^{\pi}
{\rm d}k\left[\Lambda^{-1}\left(k\right)\right]_{ab}/2\pi$. 
The `relative' \char length $\sqrt{\lao-\la_{ab}}$ directly measures the 
overlap degree of replicas $a$ and $b$. Indeed, there is a very low 
probability of finding the monomers of replicas $a$ and $b$ at positions
more distant than $\sqrt{\lao-\la_{ab}}$. Thus, the larger the 
overlap length, the less the two replicas are interacting. In the extreme 
cases: $\la_{ab}=0$, $q_{ab}\left(x,x'\right)={1\over N}
\rho_a\left(x\right)\rho_b\left(x'\right)$,
the overlap length is the same as the single chain
\char length, and replicas are not interacting at all. If instead
$\la_{ab}=\lao$, $q_{ab}\left(x,x'\right)=\rho_a\left(x\right)
\de\left(x-x'\right)$, the overlap length vanishes, and the two replicas are
interacting as strong as possible.

From now on, we will consider $\lao$ and $\la_{ab}$ as the variational
parameters\footnote
{The explicit dependence of ($\lao,\la_{ab}$) on ($\mu,\si_{ab}$)
can be worked out easily in the specific cases we are going to consider.
In the replica-symmetric case (i.e. $\si_{ab}=\si\;\forall\left(a,b\right)$)
we have $\la=\frac{\si\left(\mu+\si+2d/l^2\right)}
{\left[\left(\mu+\si\right)\left(\mu+\si+4d/l^2\right)\right]^{3/2}}$ and
$\lao=\la+\frac{1}{\left(\mu+\si \right)\left(\mu+\si+4d/l^2\right)}$.
Similar, but more complicated, equations hold for the case of one step RSB.},
with respect to which the free energy has to be stationary.
The interface breaks the translation invariance, so that the
position of the center of mass of the chain
is a relevant degree of freedom. In the following we will consider the
chain center of mass fixed in the origin, according to (\ref{5}),
with the interface moving at the variable
position $-x_0$, which then becomes another variational parameter.

Within the replica-symmetric ansatz, 
$\la_{ab}=\la\;\:\forall\left(a,b\right)$,
the quenched free energy becomes:

\beqa
\be f_q &=& -{1\over2}\left\{\ln\left({2\pi l^2\over d}\right)+
h\left[{2d\over l^2}\lao\left(1-v\right)\right] +
{2d\over l^2}\lao v h'\left[{2d\over l^2}\lao\left(1-v\right)\right]+
\right.\nonumber\\ &+& \bigl.
2\be\zeo\;\erf\left({{\tilde x}_0\over2}\right) +
\left(\be\De\right)^2\left[1-\eps\left(v,{\tilde x}_0\right)\right]\bigr\}\: ,
\label{7}
\eeqa
where ${\tilde x}_0=x_0\left(2/\lao\right)^{1/2}$, $v=\la/\lao$,
$h\left(A\right)=\sqrt{A^2+1}-A
-\ln\left({1+\sqrt{A^2+1}\over A}\right)$, $h'\left(A\right)={\rm d}h/
{\rm d}A$,
and $\eps\left(v,{\tilde x}_0\right) = \int{\rm d}x_1{\rm d}x_2
{\exp\left[-\left(x_1^2+x_2^2\right)/2\right]\over2\pi}
\;\sgn\left(x_1\sqrt{1+v}+x_2\sqrt{1-v}+{\tilde x}_0\right)$
$\sgn\left(x_1\sqrt{1+v}-x_2\sqrt{1-v}+{\tilde x}_0\right)$.
The chain is localized if the \char length $\sqrt{\lao}$ and the 
chain center of mass ${\tilde x}_0$ remain finite in the thermodynamic limit 
$\thl$. The free energy has to be minimized with respect to the variational
parameters ${\tilde x}_0$ and $\lao$, which are single-replica
quantities, and maximized with respect to $v$, which is a
two-replicas parameter \cite{MPV}.

Let us consider firstly the case of a neutral chain ($\zeo=0$).
The quenched free energy is invariant for the transformation ${\tilde x}_0
\rightarrow -{\tilde x}_0$, and it can be easily seen that the
minimum of the free energy with respect to ${\tilde x}_0$ is attained
for ${\tilde x}_{0,\rm min}=0$. Minimization with respect to $\lao$,
and maximization with respect to $v$, yield then $\lao_{,\rm min}=
l^2/\left(2d\sqrt{2v_{\rm max}-1}\right)\: ,$ where
$1/2\le v_{\rm max}\le1$ is the unique solution of the equation
$\left[\left(1+v\right)\left(2v-1\right)\right]^{1/2}
=2\be^2\De^2\left(1-v\right)^{3/2}/\pi\: .$
Thus, at any finite temperature the chain is localized, in agreement with
heuristic arguments \cite{GHLO}.

We analyze now the asymptotic behavior of our solution at high and low
temperatures. At high temperature ($\be\De\ll1$) we find:

\beq
v_{\rm max}\simeq{1\over2}+{1\over6\pi^2}\left(\be\De\right)^4
\; ;\;\lao_{,\rm 
min}\simeq{l^2\over2d}\sqrt{3}\pi\left(\be\De\right)^{-2}\: . \label{8}
\eeq
The \char length of the chain in the direction orthogonal to the
interface is $R_{\perp}=\sqrt{\lao}$. We have recovered
the scaling relation $R_{\perp}\sim\left(\be\De\right)^{-2\nu}$
coming from the Imry-Ma argument, with the exponent $\nu=1/2$
of an ideal chain. The argument is expected to fail at low temperature, 
where it is not possible to neglect anymore the role of the microscopic 
length $l$. In the limit of low temperature ($\be\De\gg1$), we indeed find:

\beqa
v_{\rm max} & \simeq & 1-\left({\pi^2\over2}\right)^{1/3}
\left({1\over\be\De}\right)^{4/3}\: ;\nonumber \\
\lao_{,\rm min} & \simeq & {l^2\over2d}\left[1+\left({\pi^2\over2}\right)^{1/3}
\left({1\over\be\De}\right)^{4/3}\right]\: .
\label{9}
\eeqa
At zero temperature, the \char length $R_{\perp}=l/\sqrt{2d}$
is of course of the same order as the microscopic length $l$.
Increasing the temperature results in a power law correction with the
non trivial exponent $4/3$.

We consider now the more general case of a non neutral chain ($\zeo\ne0$).
In this case one gets $\lao_{,\rm min}=l^2/\left(2d
\sqrt{2v_{\rm max}-1}\right)$, as in the neutral case, and the
following coupled equations for $v_{\rm max}$ and ${\tilde x}_{0,\rm min}$:

\beqa
 & & 
\left[{\left(1+v\right)\left(2v-1\right)\over\left(1-v\right)^3}\right]^{1/2}
= {2\be^2\De^2\over\pi}\exp\left(-{1\over2}{{\tilde x}_0^2\over1+v}\right)
\: ,\label{10} \\ & & {\be\De^2\over\zeo}\;
\erf\left({{\tilde x}_0\over2}\sqrt{1-v\over1+v}\right)=1\: .
\label{11}
\eeqa
There are two different regimes. The high temperature regime,
$\be\De^2/\left|\zeo\right|<1$, where equation (\ref{11}) does not admit
a solution. The variational paramaters then read $v_{\rm max}=1/2$,
$\lao_{,\rm min}=\infty$, ${\tilde x}_{0,\rm min}=
\;\sgn\left(\zeo\right)\infty$, and the chain is delocalized.
The low temperature regime, $\be\De^2/\left|\zeo\right|>1$, where a
solution exists with $v_{\rm max}>1/2$, $\lao_{,\rm min}<\infty$,
$\left|{\tilde x}_{0,\rm min}\right|<\infty$, and the chain is localized
at the interface, though the center of mass is not coincident with the
interface itself (${\tilde x}_{0,{\rm min}}\ne0$).
Therefore, lowering the temperature the chain undergoes a localization 
transition at the critical temperature $\be_c=\left|\zeo\right|/\De^2$. 
This is a second order continuous transition: if $\be\searrow\be_c$ the chain
center of mass ${\tilde x}_0$ moves continuously to infinity.

The free energy is stationary at
$v_{\rm max}\ge1/2$, both for neutral and non neutral chains, and
in the delocalized phase $v_{\rm max}=1/2$. As noticed previously, since
$v=\la/\lao\ne0$, this means that replicas are interacting in a sensible way,
even in the delocalized phase. This is in contrast with the Hartree 
approximation, which assumes instead no interaction between different 
replicas. Such approximation has been used in a new 
variational approach based on a non-stationary Green function, which
results in an asymmetric ground state even in the case of a
neutral chain \cite{SSEru}.

\vspace{10pt}

In order to ascertain the stability of the replica-symmetric solution, 
we have computed the quenched free energy within the so-called
one step RSB ansatz. We divide the $n$ replicas
in $n/m$ groups, each containing $m$ replicas, in such a way that
$\la_{ab}=\la_1$ if replicas $a$ and $b$ belong to the same group, and
$\la_{ab}=\la_2$ otherwise. Defining $v=\la_1/\lao$ and $w=\la_2/\lao$,
the quenched free energy reads:

\beqa
\be f_q & = & -{1\over2}\left\{\ln\left({2\pi l^2\over d}\right)+
{1\over m}h\left[B\right] - {1-m\over m}h\left[A\right] +
{2d\over l^2}\lao w h'\left[B\right] + \right. \nonumber \\ & + & \bigl.
2\be\zeo\;\erf\left({{\tilde x}_0\over2}\right) +
\left(\be\De\right)^2\left[1-\left(1-m\right)
\eps\left(v,{\tilde x}_0\right)
-m\eps\left(w,{\tilde x}_0\right)\right]\bigr\}\: ,
\label{12}
\eeqa
where $A=2d\lao\left(1-v\right)/ l^2$ and $B=2d\lao
\left[m\left(1-w\right)+\right.$
$\left.\left(1-m\right)\left(1-v\right)\right]/ l^2$.
This free energy has to be maximized with respect to $v$, $w$, $m$,
and minimized with respect to $\lao,{\tilde x}_0$.
Performing the change of variables $\left(v,w\right)\rightarrow
\left(s,t\right)$, with $s=v-w$ and $t=\left(1-m\right)v+mw$,
the equation $\partial f_q/\partial s=0$ has always the
solution $s=0$, which is the replica-symmetric solution, whereas 
$\partial f_q/\partial t\;|_{s=0}=0$ gives equation (\ref{10}).
The stability of this solution depends on whether it is actually a maximum
of the free energy, i.e. depends on the sign of

\beq
\left.{\partial^2f_q\over\partial s^2}\right|_{s=0}
= {m\left(1-m\right)\over 2\be}
\left\{\left({2d\over l^2}\lao\right)^2h''\left[{2d\over l^2}\lao
\left(1-t\right)\right]+
\left(\be\De\right)^2\;\eps''\left(t,{\tilde x}_0\right)\right\}\: .
\label{13}
\eeq
Notice that $h''\left[A\right]<0$ (the `entropy' contribution) and
$\eps''\left(t,{\tilde x}_0\right)=\partial^2\eps/\partial t^2>0$
(the energy contribution). At high temperature entropy is dominant,
$\partial^2f_q/\partial s^2|_{s=0}<0$,
and $s=0$ is in fact a maximum. But at low temperature the energy
contribution prevails, $\partial^2f_q/\partial s^2|_{s=0}>0$ and
$s=0$ becomes a minimum, signalling that the replica-symmetric solution
is unstable. Therefore, within the Gaussian variational approach,
we find a glass transition to a phase characterized by the presence
of many metastable states separated by high energy barriers, according
to the usual interpretation of RSB \cite{MPV}.
If the transition takes place in the way sketched before, i.e. if it
is continuous, the transition temperature is given by the equation
$\partial^2f_q/\partial s^2|_{s=0}=0$.
In principle, however, it could be possible to have a discontinuous
transition if another local maximum is present which provides a better
extremum for the free energy when the replica-symmetric solution
is still stable.

In the case of a neutral chain ($\ze_0=0,{\tilde x}_0=0$), we can easily
compute the temperature
$\be_t$ at which the replica-symmetric solution becomes unstable.
The solution of equations $\partial^2f_q/\partial s^2|_{s=0}=0$ and
$\partial f_q/\partial t\;|_{s=0}=0$ yields $v_t=1/\sqrt{2}$,
$\be_t^2=\pi\sqrt{2+5/\sqrt{2}}/\De^2$, and $\lao_{,t}=\left(l^2/2d\right)
\sqrt{1+\sqrt{2}}$ for the square of the \char length at the transition.

The occurrence of RSB in such a simple
model, i.e. in a model without any self-interaction between different
monomers, is higly surprising and counter-intuitive.
In fact, the energy term $-\be\sum_{i=0}^N\zei\;\sgn\left(x_i\right)$
alone does not provide any frustration. Each monomer simply chooses
its preferred side and frustation arises from the competiton between energy
and chain connectivity, which is an entropic effect. Thus, at
sufficiently low temperature we would expect no frustration at all.

\vspace{10pt}

In conclusion, we have studied the localization of an ideal copolymer
chain at a selective interface. Within the Gaussian variational
method and a replica-symmetric ansatz, an explicit calculation of the
relevant physical quantities has been possible.
At high temperature the results agree with the predictions based on
Imry-Ma argument \cite{GHLO,SoDa}, whereas at low temperature a non
trivial dependence on the microscopic bond length is present.
Furthermore, the replica-symmetric solution has been shown to be unstable
within a one step RSB scheme. This would imply a breaking of the ergodicity
of the system, in contrast with simple intuitive arguments.
The Gaussian variational approach has been thought to be
effective in describing correctly the physics of disordered frustrated
systems \cite{ShaGut2,MePa,MKD,Eng}. We deem therefore a quite interesting 
issue to try to evidentiate RSB with Monte Carlo simulations, or
to understand why the Gaussian variational method fails in predicting it,
if this should be the case.

\vspace{10pt}

We thank Flavio Seno for introducing us to this interesting
subject, and Jort van Mourik, Jayanth Banavar and Alain Barrat
for useful discussions.



\end{document}